\documentclass[english,twocolumn,showpacs,preprintnumbers,nofootinbib]{revtex4}
\usepackage[T1]{fontenc}
\usepackage[latin1]{inputenc}
\usepackage{amsmath}
\usepackage{color}
\usepackage{graphicx}

\makeatletter

\makeatletter
\usepackage{epstopdf}

\def\lsim{\mathrel{\rlap{\lower4pt\hbox{\hskip1pt$\sim$}}
    \raise1pt\hbox{$<$}}}
\def\gsim{\mathrel{\rlap{\lower4pt\hbox{\hskip1pt$\sim$}}
    \raise1pt\hbox{$>$}}}
\def\sqr#1#2{{\vcenter{\vbox{\hrule height.#2pt
         \hbox{\vrule width.#2pt height#1pt \kern#1pt
         \vrule width.#2pt}
         \hrule height.#2pt}}}}

\makeatother

\usepackage{babel}
\makeatother
\begin{document}

\title{Dark Energy-Dark Matter Interaction and putative violation of the
Equivalence Principle from the Abell Cluster A586}

\author{O. Bertolami}

\thanks{Also at Centro de F\'{\i}sica dos Plasmas, IST}

\email{orfeu@cosmos.ist.utl.pt}

\author{F. Gil Pedro}

\email{fgpedro@fisica.ist.utl.pt}

\affiliation{Departamento de F\'{\i}sica, Instituto Superior T\'{e}cnico \\
 Av. Rovisco Pais 1, 1049-001 Lisboa, Portugal}

\author{M. Le Delliou}

\email{delliou@cii.fc.ul.pt}

\affiliation{Centro de F\'{\i}sica Te\'{o}rica e Computacional, Universidade
de Lisboa \\
 Av. Gama Pinto 2, 1649-003 Lisboa, Portugal}

\begin{abstract}
We show that the Abell Cluster A586 exhibits evidence of the interaction
between dark matter and dark energy and argue that this interaction
implies a violation of the Equivalence Principle. This violation is
found in the context of two different models of dark energy-dark matter
interaction. We also argue, based on the spherical symmetry of the
Abell Cluster A586 that skewness is not the most general quantity
to test the Equivalence Principle.
\end{abstract}

\pacs{98.80.-k,98.80.Cq\hspace{4cm} Preprint DF/IST-3.2007}

\maketitle

\textit{Introduction.} It has become rather consensual that the problem of the nature 
of dark energy and dark matter (hereafter DE
and DM, respectively) is crucial in contemporary cosmology. Even though, observational data is fully consistent 
with the $\Lambda$CDM parametrization, in order to get a deeper insight into the nature of 
DE and DM one must consider more complex models and, in particular, the interaction of those components.
However, so far no observational evidence of this interaction has
been presented. In this work, we argue that study of the Abell Cluster
A586 exhibits evidence of the interaction between DE and DM. Furthermore,
we show that this interaction implies a violation of the Equivalence
Principle (EP). Our results are obtained in the context of two distinct
phenomenologically viable models for the DE-DM interaction. We consider
the generalized Chaplygin gas (GCG) model \cite{Bento02}, a unified
description of DE and DM, where interaction is an automatic feature
of this proposal, but also a less constrained approach where DE and
DM are regarded as two independent components, but interacting (see
e.g. \cite{Amendola}). We show that interaction between DE and DM
implies a violation of the EP between DM and baryons which can be
detected in self-gravitating systems in stationary equilibrium. For
sure, the EP -- that is, the universality
of free fall -- is one of the cornerstones of general relativity,
however its validity at cosmological scales has never been directly
tested (see \cite{Bertolami06} and references therein). The EP can
be expressed in terms of the bias parameter, $b$, defined as ratio
of baryon over DM density, at a typical clustering scale (Mpc). Should
the EP hold, $b$ would be a constant over time since then all clustering
species would fall equivalently under the action of gravity. Inversely,
clustering should reflect the violation of the EP through a different
behaviour for both species. Interaction between DM and DE induces a
time evolution of $b$.


In this work we shall focus on the effect of interaction on clustering as revealed by the
Layzer-Irvine equation. Given that the EP concerns the way matter falls in the
gravitational field, considering the clustering of matter against the cosmic expansion and
the interaction with DE seems to be a logical way to test its validity. In what follows we
shall see that DE-DM interaction implies a departure from virial equilibrium. First, we
will set the formalism to address the DE-DM interaction and consider two
phenomenologically viable models: one based on a \textit{ad hoc} DE-DM interaction
\cite{Amendola}, the other in the GCG with an explicit identification of the DE and DM
components \cite{Bento04}. Our observational inferences are based on the Abell Cluster
A586 given its stationarity, spherical symmetry and wealth of available observations
\cite{Cypriano:2005}.


\textit{Quintessence model with DE-DM interaction.} The Bianchi identities
with coupling $\zeta$ give origin to the following homogeneous energy
conservation equations: \begin{equation}
\dot{\rho}_{DM}+3H\rho_{DM}=\zeta H\rho_{DM}\:,\label{eq:DMcons}\end{equation}
 \begin{equation}
\dot{\rho}_{DE}+3H\rho_{DE}(1+\omega_{DE})=-\zeta H\rho_{DM}\:.\label{eq:DEcons}\end{equation}
Notice that these equations imply that the energy exchange between DE and DM is adiabatic (see e.g. \cite{Pavon04} 
and references therein). Moreover, the basic assumptions used in these equations are a constant equation
of state parameter $p_{DE}=\omega_{DE}\rho_{DE}$ and the following
scaling with respect DM energy density \begin{equation}
\frac{\rho_{DE}}{\rho_{DM}}=\frac{\Omega_{DE_{0}}}{\Omega_{DM_{0}}}a^{\eta}\:,\label{eq:DEsDM}\end{equation}
 for a constant $\eta$. From the time derivative of Eq. (\ref{eq:DEsDM})
inserted into Eq. (\ref{eq:DEcons}) together with Eq. (\ref{eq:DMcons})
yields: \begin{equation}
\zeta=-\frac{(\eta+3\omega_{DE})\Omega_{DE_{0}}}{\Omega_{DE_{0}}+\Omega_{DM_{0}}a^{-\eta}}\:.\label{eq:DefZeta}\end{equation}
 The solution of Eq. (\ref{eq:DMcons}) is given by \begin{eqnarray}
\rho_{DM} & = & a^{-3}\rho_{DM_{0}}e^{\int_{1}^{a}\zeta\frac{da}{a}}\nonumber \\
 & = & a^{-3}\rho_{DM_{0}}\left[\Omega_{DE_{0}}a^{\eta}+\Omega_{DM_{0}}\right]^{-\frac{(\eta+3\omega_{DE})}{\eta}}\:.\label{eq:rhoDM}\end{eqnarray}
 The DE evolution is then derived from the scaling directly, or from
Eq. (\ref{eq:DEcons}) combined with the scaling: \begin{eqnarray}
\rho_{DE} & = & a^{\eta-3}\rho_{DE_{0}}e^{\int_{1}^{a}\zeta\frac{da}{a}}\nonumber \\
 & = & a^{\eta-3}\rho_{DE_{0}}\left[\Omega_{DE_{0}}a^{\eta}+\Omega_{DM_{0}}\right]^{-\frac{(\eta+3\omega_{DE})}{\eta}}\:.\label{eq:rhoDE}\end{eqnarray}
In this model, from Eq. (\ref{eq:rhoDM}) one can see that the bias
parameter depends on time as follows: \begin{equation}
b={\frac{\rho_{B}}{\rho_{DM}}}=\frac{\Omega_{B_{0}}}{\Omega_{DM_{0}}}\left[\Omega_{DE_{0}}a^{\eta}+\Omega_{DM_{0}}\right]^{\frac{(\eta+3\omega_{DE})}{\eta}}\:.\label{bias2}\end{equation}

\textit{The GCG model.} Let us now consider the GCG model with an
explicit identification of DE and DM, as discussed in Ref. \cite{Bento04}.
The GCG model is considered here as it fares quite well when confronted
with various phenomenological tests: high precision Cosmic Microwave
Background Radiation data \cite{Bento3}, supernova data \cite{Supern,Bertolami1,Bento4},
gravitational lensing \cite{Silva}, gamma-ray bursts \cite{Bertolami2},
cosmic topology \cite{Bento5} and time variation of the electromagnetic
coupling \cite{Bento07}. In order to obtain a suitable structure
formation behaviour at linear approximation, $\omega_{DE}=-1$ (see
\cite{Bento04} and references therein). For the GCG admixture of
DE and DM, the equation of state is given by \cite{Bento02}:
\begin{equation}
p=-\frac{A}{\rho^{\alpha}}\:,\label{GCG}\end{equation}
 where $A$ and $\alpha$ are positive constants. From \cite{Bento04},
the DM and DE expressions for a constant DE equation of state are
given by\begin{align}
\rho_{DM}= & \rho_{DM_{0}}a^{-3(1+\alpha)}\left(\frac{\Omega_{DE_{0}}+\Omega_{DM_{0}}}{\Omega_{DE_{0}}+\Omega_{DM_{0}}a^{-3(1+\alpha)}}\right)^{\frac{\alpha}{1+\alpha}}\:,\label{GCGDM}\\
\rho_{DE}= & \rho_{DE_{0}}\left(\frac{\Omega_{DE_{0}}+\Omega_{DM_{0}}}{\Omega_{DE_{0}}+\Omega_{DM_{0}}a^{-3(1+\alpha)}}\right)^{\frac{\alpha}{1+\alpha}}\:,\label{GCGDE}\end{align}
so that we recover Eq. (\ref{eq:DEsDM}), but now with $\eta=3(1+\alpha)$
and $\omega_{DE}=-1$.

\textit{The Generalized Layzer-Irvine equation.} Let us now turn to
the cosmological gravitational collapse and its implication for the
EP. The core of our approach lies on deviation of the classical virial
equilibrium, in its standard Layzer-Irvine equation form. We argue
that A586 data allows to establish this departure independently of
the DE-DM interaction model considered. It is possible to write the
energy density conservation for non-relativistic self-gravitating
dust-like particles through the so-called Layzer-Irvine equation \cite{Peebles}.
The kernel of the method is to consider the Newtonian kinetic energy,
$K$, per unit mass, while keeping the average momentum and mass,
$M$, constant: \begin{equation}
MK=\frac{1}{2a^{2}}\left\langle \frac{p^{2}}{m}\right\rangle \propto a^{-2}\:,\label{Kinetic}\end{equation}
 where $a$ is the scale factor of the Robertson-Walker metric. It
then follows that: \begin{equation}
\rho_{K}\equiv MdK/dV=d(MK)/dV\propto a^{-2}\:.\label{Varkinetic}\end{equation}
 It is assumed that the mass evolution of the cluster remains constant
over the course of the observation. The energy transfer between DM and DE is negligible at
this point.The potential energy per unit mass derives from the definition of the
auto-correlation function, $\xi(r)$, \cite{Peebles}\begin{align} W & =-2\pi
Ga^{2}\rho_{DM}\int dr\xi(r)r\:,\label{correlationf}\end{align}
 where we have replaced the background energy density by the DM energy
density. After considering the DE-DM interaction, it follows that
\begin{equation}    
W\propto a^{2+d ln\rho_DM / d ln a}=a^{\zeta-1}\:.\label{W}\end{equation}
and hence \begin{equation}
\rho_{W}\equiv MdW/dV=d(MW)/dV\propto a^{\zeta-1}\:.\label{DW}\end{equation}
 This is the source of difference from the usual dust case. The Layzer-Irvine
equation for the energies per unit volume is just a chain rule of
time derivative for the energy density where the time is parameterized
by the scale factor, hence: \begin{equation}
\frac{d}{dt}\left(\rho_{DM}\right)=\dot{a}\frac{\partial}{\partial a}\left(\rho_{DM}\right)=-\left[2\rho_{K}+(1-\zeta)\rho_{W}\right]H\:,\label{rhoDM}\end{equation}
 from which follows \begin{equation}
\dot{\rho}_{DM}+\left(2\rho_{K}+\rho_{W}\right)H=\zeta\rho_{W}H\:,\label{eq:CoupledL-I}\end{equation}
 where $H=\dot{a}/a$ is the expansion rate.

Furthermore, writing in terms of the virial equilibrium factor $2\rho_{K}+\rho_{W}$
and the departure to static equilibrium, due to the DE-DM interaction,
Eq. (\ref{eq:CoupledL-I}) becomes \begin{equation}
\dot{\rho}_{DM}+H(2\rho_{K}+\rho_{W})=-\frac{(\eta+3\omega_{DE})H}{1+\Omega_{DM_{0}}/\Omega_{DE_{0}}a^{-\eta}}\rho_{W}\:.\label{equilibrium}\end{equation}
 As before, it is possible to see from the equivalent of Eq. (\ref{eq:DEsDM})
for the GCG model (for which $\omega_{DE}=-1$ \cite{Bento04}) that
one can map Eq. (\ref{equilibrium}) for the generic interaction model
into the GCG model via the relationship $\eta=3(1+\alpha)$. 
Next we shall apply these equations to the stationary Abell Cluster A586 for which
$\rho_{K}$ and $\rho_{W}$ can be computed, so as to compare with the observed local
measurements with the homogeneous-spawned interaction term: \begin{equation}
2\rho_{K}+\rho_{W}=\zeta\rho_{W}\:.\label{comparison}\end{equation}

\textit{The Abell Cluster A586.} In order to estimate the coupling
between DE and DM from Eq. (\ref{comparison}) one has to find a particular
cluster to compute $\rho_{K}$ and $\rho_{W}$. It is convenient that
the cluster is as spherical as possible and close to stationary equilibrium.
Under these conditions, one can approximate the kinetic and potential
energy densities as: \begin{align}
\rho_{K}= & M\frac{d}{dV}K\simeq M\frac{K}{V}\simeq\frac{9}{8\pi}\frac{M_{Cluster}}{R_{Cluster}^{3}}\sigma_{v}^{2}\label{eq:rhoK}\\
\rho_{W}= & M\frac{d}{dV}W\simeq M\frac{W}{V}\simeq-\frac{3}{8\pi}\frac{G}{<R>}\frac{M_{Cluster}^{2}}{R_{Cluster}^{3}}\:,\label{eq:rhoW}\end{align}
 where $M_{Cluster}$ and $R_{Cluster}$ are the cluster's total mass
(galaxies, DM and intra-cluster gas) and radius, $\sigma_{v}$ is
the velocity dispersion as determined globally from weak lensing,
and $<R>$ is the mean intergalactic distance \cite{Cypriano:2005}.

The cluster must be also relaxed, since the core of our method consists
in estimating the EP violation from a deviation from the standard
form of the cosmic virial theorem defined by Eq. (\ref{eq:CoupledL-I})
set with no interaction.

Given these constraints a particularly suitable cluster for our purpose
is the Abell cluster A586 \cite{Cypriano:2005}. It is found that
the mass profile in this particular cluster is approximately spherical
and that it is a relaxed cluster, since it has not undergone any important
merging process in the last few Gyrs \cite{Cypriano:2005}. The agreement
between dynamical (velocity dispersion and X-ray) and non-dynamical
mass estimates (weak-lensing) indicates that A586 is in fact a relaxed
cluster.

Considering that gravitational weak lensing is independent from equilibrium
assumptions about the dynamical state of the cluster, it turns out
to be the best mass estimator. Therefore, in our analysis we assume
\cite{Cypriano:2005}: \begin{equation}
M_{Cluster}=(4.3\pm0.7)\times10^{14}\: M_{\odot}\label{eq:MCluster}\end{equation}
 which corresponds to the total mass inside a 422 kpc radius region
estimated using gravitational weak lensing.

In order to have a coherent set of data, we take for the velocity
dispersion \cite{Cypriano:2005}: \begin{equation}
\sigma_{v}=(1243\pm58)\: kms^{-1}\label{eq:sigmav}\end{equation}
 as computed from gravitational weak lensing measurements.

The mean intergalactic distance is estimated using the coordinates
(right ascension-$\alpha_{c}$ and declination-$\delta_{c}$) of the
31 galaxy sample provided in Ref. \cite{Cypriano:2005}. Given that
weak gravitational lensing data concerns a 422 kpc radius spherical
region and the 31 galaxies lie within a $570h_{70}^{-1}~kpc$ region,
one has to select from the original sample the galaxies that lie within
the range of interest. Since at the cluster's distance, one arcsecond
corresponds to 2.9 kpc, we select from our sample the galaxies that
have $\alpha_{c}$ and $\delta_{c}$ such that: \begin{equation}
\sqrt{(\alpha_{c}-\alpha_{center})^{2}+(\delta_{c}-\delta_{center})^{2}}\leq\Delta_{max}\:,\label{samplingcond}\end{equation}
 where $\alpha_{center}$ and $\delta_{center}$ are the coordinates
of the center of the cluster and $\Delta_{max}=145''$ is the angular
dimension corresponding to a radius of 422 kpc. From this procedure,
we build a sub-sample containing 25 galaxies. From this sub-sample
coordinates one can estimate the mean intergalactic distance by elementary
trigonometry, the distance between any two galaxies $i$ and $j$
with coordinates $(\alpha_{ci},\delta_{ci})$ and $(\alpha_{cj},\delta_{cj})$
is given by $r_{ij}^{2}=2d^{2}\left[1-cos(\alpha_{ci}-\alpha_{cj})cos\delta_{ci}cos\delta_{cj}-sin\delta_{ci}sin\delta_{cj}\right]$,
where $d$ is the radial distance from the center of the cluster to
Earth. Therefore the mean intergalactic distance $<R>$ is \begin{equation}
<R>=\frac{2}{N_{gal}(N_{gal}-1)}\sum_{i=1}^{N_{gal}}{\sum_{j=1}^{i}{r_{ij}}}\:,\label{eq:Rmed}\end{equation}
 where $N_{gal}$ is the number of galaxies in the sample. In our
sub-sample, $N_{gal}=25$ and hence we get the estimate $<R>=309$
kpc. Using Eqs. (\ref{eq:MCluster}), (\ref{eq:sigmav}) and $<R>$
we can estimate the kinetic and potential energy densities, Eqs. (\ref{eq:rhoK})
and (\ref{eq:rhoW}): \begin{align}
\rho_{K}= & (2.14\pm0.55)\times10^{-10}Jm^{-3}\:,\label{eq:rhoKN}\\
\rho_{W}= & (-2.83\pm0.92)\times10^{-10}Jm^{-3}\:,\label{eq:rhoWN}\end{align}
 where the errors were computed using linear error propagation.

It is worth mentioning that \begin{equation}
\frac{\rho_{K}}{\rho_{W}}\simeq-0.76\pm0.05\:,\label{ratioKW}\end{equation}
 instead of $-0.5$ as one would expect for a relaxed cluster considering
the standard form of the cosmic virial theorem and no DE-DM interaction.

\textit{DE-DM interaction and putative evidence of violation of the
EP.} In what follows we use our estimates of $\rho_{K}$ and $\rho_{W}$,
Eqs. (\ref{eq:rhoKN}), (\ref{eq:rhoWN}), and the latest cosmological
WMAP data \cite{Spergel:2006hy} to show the evidence of DE-DM interaction.
We also demonstrate that this interaction implies a violation of the
EP between DM and baryons.

Let us first look at the quintessence model with DE-DM interaction.
From Eqs. (\ref{eq:DMcons}) and (\ref{eq:DEcons}) the DE-DM interaction
is exhibited through a non-vanishing $\zeta$ or equivalently, from
Eq. (\ref{eq:DefZeta}), by the condition $\eta\neq-3\omega_{DE}$.

Thus, assuming that $\omega_{DE}=-1$, $\Omega_{DE_{0}}=0.72$, $\Omega_{DM_{0}}=0.24$,
one can estimate $\eta$ for which Eq. (\ref{comparison}) is satisfied
for the redshift of the A586, $z=0.1708$. We find that: \begin{equation}
\eta=3.82_{-0.17}^{+0.18}\:.\label{eq:etaN}\end{equation}

Thus, since Eq. (\ref{eq:etaN}) satisfies the condition $\eta\neq-3\omega_{DE}$,
one concludes that DE and DM are interacting. Notice that, as observations
suggest a recent DE dominance, then $\zeta<0$, and from there follows
that $\eta>-3\omega_{DE}$. This means that Eq. (\ref{eq:etaN}) not
only suggests that DE and DM are interacting, but also, as expected,
that the energy transfer flow is from DM to DE.

Let us now turn to the CGC model. With the identification of components
suggested in \cite{Bento04}, DE-DM interaction is expressed by the
condition $\alpha\neq0$. In order to see the effect of interaction
in the GCG model, we proceed as before using Eqs. (\ref{comparison}),
(\ref{eq:rhoKN}) and (\ref{eq:rhoWN}), from which follows: \begin{equation}
\alpha=0.27_{-0.06}^{+0.06}\:.\label{eq:alpha}\end{equation}

Thus, the condition $\alpha\neq0$ holds, meaning that the A586 data
is consistent with the identification of components suggested in \cite{Bento04}
for the CGC model. Notice that for $\alpha=0$ the GCG model corresponds
to the $\Lambda$CDM model. Moreover, it is interesting to point out
that the value $\alpha\sim0.27$ is approximately consistent with
values found to match the bias and its growth from the 2dF survey
(see \cite{Bento04} and references therein).

\begin{figure}[h!]
\centering \includegraphics[width=0.34\textwidth]{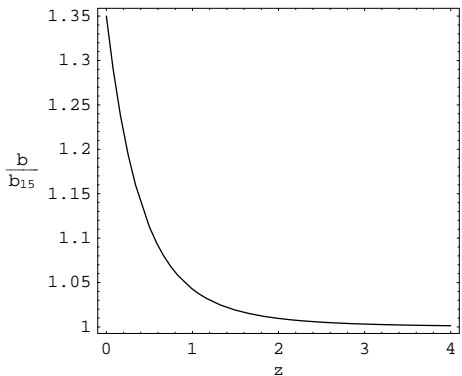}

\caption{Normalized gravitationally induced bias parameter as a function of the redshift,
where $b_{15}\equiv b(z=15)$, $z=15$ being a typical condensation time.}

\label{fig:bias}
\end{figure}

Evidence on a possible violation of the EP implies the time dependence
of the bias parameter. We depict in Fig. \ref{fig:bias}, the evolution
with redshift of the normalized bias parameter predicted by Eq. (\ref{bias2})
where only gravitational effects are considered. Even though other astrophysical
effects might affect the way DM and baryons fall under gravity, 
for EP purposes, gravity is the only relevant interaction
that offers a clear drift on a cosmological time scale. Clearly, one
expects that for large samples those effects would average out for
non-cosmological drifts and thus lead to possible detection in large
surveys.

%


Figure \ref{fig:bias} shows that $b(z)/b_{15}$ has undergone a sharp
change in the recent past, a clear signal of the violation of the
EP due to the DE-DM interaction. This abrupt variation corresponds
to the period when energy transfer from DM to DE becomes significant
($z\sim0.5$).

\textit{Discussion and Conclusions.} In this work we have argued that
the properties of the A586 suggest evidence of the interaction between
DE and DM. We stress that the considered models to describe the DE-DM
interaction are consistent with known phenomenological constraints.
We have also argued that this interaction does suggest a violation
of the EP that should be detectable in large scale cluster surveys
via the analysis of the time dependence of the bias parameter. We
find that this violation is independent of the interaction model between
DE and DM and entails a redshift evolution of bias parameter given
by Eq. (\ref{bias2}) and depicted in Figure \ref{fig:bias}. Our
conclusions are independent of the DE-DM interaction model, generic
or GCG. Actually, a violation of the EP is reported to be found in
other DE models \cite{Alimi06}. For the GCG model we find that the
detection of interaction precludes the $\Lambda$CDM model ($\alpha=0$).
Furthermore, the obtained value for $\alpha$ is approximately consistent
with results for the bias and its growth obtained by the 2dF survey
\cite{Bento04}. Consistency of our results with observational data concerning interaction 
\cite{Amendola07} and further implications of the detected interaction between 
DE and DM, for instance, in what concerns the motion of the satellite Sagittarius galaxy 
\cite{Kesden}, are discussed in \cite{OBPedro07b}. 

It is interesting to point out that our results indicate evidence
for violation of the EP between baryons and DM using data extracted
from the A586, a notoriously relaxed and spherically symmetric structure.
This seems to imply that the suggestion that cosmological evidence
for a violation could be detected via skewness \cite{AmendolaQuercellini04}
does not hold. Indeed, spherical symmetry implies that skewness vanishes
given that it is an odd parity spatial function. Thus, while the virial
equilibrium may in principle reveal the violation of EP due to the
DM-DE interaction, skewness is unable, by definition, to detect it
in this particular symmetry. The spherical symmetry of A586 and our
detection of violation of the EP via virial equilibrium exemplifies
this point.

\textit{Acknowledgments.} The work of MLeD is supported by FCT (Portugal)
under the grant SFRH/BD/16630/2004 and hosted by J.P.Mimoso and CFTC,
Lisbon University. The work of OB is partially supported by the FCT
project POCI/FIS/56093/2004. The authors would like to thank Catarina
Lobo for the information on clusters.

\end{document}